\documentstyle{article}
\title{\sc RADIATION FIELD ON SUPERSPACE}
\author{ Pedro F. Gonz\'alez-D\'{\i}az.\\
Instituto de Matem\'aticas y F\'{\i}sica Fundamental\\
Consejo Superior de Investigaciones Cientificas\\
Serrano 121, 28006 Madrid (SPAIN)\\
}
\date{March 18, 1994}
\begin{document}
\maketitle
\large
\setlength{\baselineskip}{0.5cm}
\vspace{3cm}

We study the dynamics of multiwormhole configurations within the
framework of the Euclidean Polyakov approach to string theory,
incorporating a modification to the Hamiltonian which
makes it impossible to interpret the Coleman
$\alpha$-parameters of the effective interactions as a quantum field
on superspace, reducible to an infinite tower of fields on space-time.
We obtain a Planckian probability measure for the $\alpha$ that allows
$\frac{1}{2}\alpha^{2}$ to be interpreted as the energy of the quanta of
a radiation field on superspace whose values may fix the coupling constants.


\pagebreak

\section{CAN WORMHOLES FIX THE COUPLING CONSTANTS?}

Discussion on cosmological objects such as wormholes in spacetime within
the realm of Euclidean quantum gravity has been often shadowed by the
fact that general relativity is not renormalizable. The very
tempting possibility first suggested by Coleman [1] that the
values of the coupling constants can be predicted from first
principles by using wormhole interaction remains at present
inconclusive due mostly to the shortcomings of the underlying
formulation of quantum gravity. Originally, Coleman's idea
was that topological
fluctuations of space-time can be regarded as point interactions
at which particles appeared and disappeared on the low-energy
asymptotic region, and therefore, described by effective
interaction terms $\theta_{i}$. In dilute wormhole approximation [2],
such wormhole can be represented as bi-local effective
addictions to the action, resulting finally in a path
integral for the localising Coleman parameters $\alpha$,
with probability measure [3] $\mu(\alpha)=P(\alpha)Z(\alpha)$,
where $P(\alpha)=e^{-\frac{1}{2}D\alpha^{2}}$, $D\propto e^{I}$,
with $I$ the Euclidean action, and
\[Z(\alpha)=\int d[\phi]e^{-\int
d^{4}xg^{\frac{1}{2}}(L+\alpha_{j}\theta_{j})}\]
is the path integral for the parent universe, with $\phi$ a scalar
field and $L$ the Lagrangian.

Using the saddle-point approximation for a single sphere and an
effective action $\Gamma=-\frac{3}{8G^{2}\Lambda(\alpha)}+f(\alpha)+\Lambda
g(\alpha)+...$,
it was obtained [1,3] $Z(\alpha)=e^{\Gamma}$ or $e^{e^{-\Gamma}}$.
Once a most probable vanishing cosmological constant $\Lambda$
is fixed [1] through the first term of $\Gamma$, one would then
expect that $Z(\alpha)$ concentrated at the minimum of $f(\alpha)$
for a given $\alpha$, so fixing as well the value of the coupling
constants involved in $L$ [1-4].

Apart of general problems such as nonrenormalizability (or
nonunitarity in $R^{2}$-theories) and conformal divergences
of Euclidean quantum gravity [5], the Coleman mechanism
showed some important, more specific problems: (i) it led
the pion mass down to zero and the free fermion mass up to
the cut off scale [2], (ii) giant wormholes would be
produced with a high density [6], and (iii) parameters $\alpha$
were shown to depend on an arbitrary cut off which is
to be introduced to prevent divergence of the probability
distribution [3] (The alleged Polchinski's phase problem [7],
if it existed, is harmless for $Z(\alpha)=e^{-\Gamma}$.)

Problem (i) can however be overcome by invoking [8] a further
level of quantization that arises from using wormholes in
a mixed quantum state, rather than a pure state, and finally
leads to identifying perturbations to the effective action
with minus the entropy of the system. As to difficulty (ii),
Coleman and Lee [9] have proposed a mechanism for stopping
giant wormholes from forming, but even if such a mechanism
failed, one should notice that the largest possible scales
for wormholes which are solution to higher-derivative
gravity theories coupled to an axionic field, still are
submicroscopic [10,11]. It appears then that out of
the specific problems originally posed, it is only (iii)
that still remains.

Concerning the fact that general relativity is not
renormalizable (or nonunitary), and the Euclidean action
is not positive definite, it was thought [12-14] that
these problems could be circumvented by working in the
framework of the Euclidean Polyakov approach to string
theory.
Multiwormhole configurations in Polyakov string theory have been studied
by looking at the wormholes as the handles on a Riemann surface of
genus $g$, with $g$ giving the number of handles or wormholes in the
configuration [12]. The Green function that describes the effects of such
wormholes on first order tachyonic amplitudes was calculated by
Lyons and Hawking [14] as a path integral over all space-time coordinates
$x_{\mu}$ on the Riemann surface, with all the fields having the
same values at the points on the two circles which result after
cutting the handles in such a way that they become divided
in two topologically separated parts. In this case, the points were
identified by the projective transformations of the Schottky group
on each pair of circles, and the Green function can be written as [12-14]
\begin{equation}
<x_{\mu}(z_{1}).x_{\nu}(z_{2})...>=\int
d[x_{\mu}]x_{\mu}(z_{1})x_{\nu}(z_{2})...\prod_{r}\prod_{n}\delta(x_{n}-x'_{n})e^{-I},
\end{equation}
where $I$ is the Euclidean action, $r$ runs from 1 to $g$,
\begin{equation}
x=\sum x_{n}e^{i\zeta n},
\end{equation}
and the delta function ensures that $\zeta$ on one circle is identified
with $\zeta '$ on another. On can express the Green function (1) in
terms of the handle quantum state on the circles
using the Fourier transform of the delta
function for the zero mode, and expanding the delta function for the
nonzero modes in terms of the complete set of orthonormal harmonic-oscillator
eigenstates which are the solutions of the string analogue
of the Wheeler DeWitt equation [14]
\begin{equation}
H_{WDW}\Psi_{nm_{n}^{(i)}}=[-\frac{\partial^{2}}{\partial
x_{0}^{2}}+\frac{1}{2}\sum_{n>0,i}(-\frac{\partial^{2}}{\partial(Y_{n}^{(i)})^{2}}+n^{2}(Y_{n}^{(i)})^{2})]\Psi_{nm_{n}^{(i)}}=0,
\end{equation}
with solution ($i=1,2$)
\begin{equation}
\Psi_{nm_{n}^{(i)}}\propto
e^{-\frac{1}{2}n(Y_{n}^{(i)})^{2}}H_{m_{n}^{(i)}}(n^{\frac{1}{2}}Y_{n}^{(i)})\Psi_{K}(x_{0}), \Psi_{K}(x_{0})=e^{iK.x_{0}}
\end{equation}
(3) is the canonically quantised version of the Hamiltonian constraint
derived [14] from the string Euclidean action for the field $x$
on the region of the complex plane outside a disc of radius
$r=-\frac{\ln\mid z\mid}{t}$ ($t$ is some Euclidean time), $K$
is the momentum of the zero mode $n=0$, and
\[Y_{n}^{(1)}=\frac{1}{2}(x_{n}+x_{-n}),
Y_{n}^{(2)}=\frac{1}{2i}(x_{n}-x_{-n}).\]

Now, as in the 4-dimensional case [15], one can calculate the
effect on Green functions in the fundamental region, by
doing a path integral over all fields $x^{\mu}$ on the
complex plane with the boundary conditions on the circles
given by a set of values $Y_{n}^{(i)}$, weighting (filtering)
with the wave function $\Psi_{nm_{n}}^{(i)}$. Again as in
the space-time wormhole case, the effect will be given
by a vertex operator located at the center of the circles.
One can see that, after integrating over the fields, the
resulting path integral contains a factor $\kappa=(K^{2}+\sum\mid n\mid
m_{n}^{(i)}-2)^{-1}$.
We obtain thus a bi-local effective action [12]
\[-\int d\sigma_{1}\int d\sigma_{2}\sum\int d^{4}K\kappa
V_{p}(\sigma_{1})V_{p}(\sigma_{2}),\]
where the $V_{p}$ are the handle vertex operators for
the fields on the two circles. One can again convert the
bi-local action to a local action, by introducing $\alpha$
parameters, to finally obtain a probability measure with the
same general form as for the 4-dimensional case [3]. Nevertheless,
since now the $\alpha$ parameters are labelled by momentum $K$,
and also the occupation numbers $m_{n}^{(i)}$ for modes $n>0$,
these parameters
should be interpreted as a quantum field on the infinite
dimensional superspace of all coordinate field values on a single circle.
Such a quantum field would be regarded  [12,13] as an infinite tower of fields
on
the usual space-time minisuperspace; each stage along this tower is
labelled by the quantum number $m_{n}^{(i)}$ that defines the excited states of
the
basis set of solutions (4). In light of this interpretation, it was
concluded [12,13] that the $\alpha$'s could not be regarded as a set of
coupling
constants to be fixed by quantum measurement. This new serious
problem, together with the above-mentioned
fact that
the value of $\alpha$ on which the probability distribution is concentrated
will depend on the arbitrary choice of the cut off required to make the
distribution convergent, becomes the main difficulty for the
wormhole Big Fix. However, by using a modified, more consistent
general string-theory approach, we shall show in what follows
that these two difficulties should not actually occur.

\section{STRINGY WORMHOLE INTERACTION}

It is not quite clear that the conclusion obtained in the
precedent section for stringy wormholes
can be maintained,
since the picture underlying it fails to meet two main requirements of
string theory. First, there is the fact that in string theory the
Hamiltonian should be modified [16] by the addition of an (infinite) constant,
in such a way that just the ground state of the harmonic oscillators,
multiplied by the wave function of the zero mode, will obey the
Hamiltonian constraint, so that the Virasoro condition will be $K^{2}=2$,
which is in contradiction with the general states required to generate
an infinite tower of field on space-time, and hence with the use of
a full $\delta$-function for all the modes in the Green function. On
the other hand, the model does not account, but indeed contradicts as
well the known string-theoretic demand of a finite maximum resolution [17,18]
which could erase a great deal of the microscopic detail of the above
formalism. Using just simply connected handles which, under cutting,
produce mutually disconnected discs, is an assumption that leads
to ambiguity in the definition of the quantum state on the discs,
while still unnecessarily restricting to pure states [19]. If we
kept a full $\delta$ function in (1), then one could actually
choose any set of orthogonal eigenstates other than (4) for the
$\delta$ function of the $n>0$ modes [14]. This freedom in the
choice of a pure quantum state would ultimately correspond to
an ambiguity in the cut off that makes the probability convergent
in space-time wormholes.

In the present paper we shall follow a different approach where these
shortcomings are avoided, and leads to an $\alpha$-parameters
interpretation similar to that by Coleman [1] and Preskill [4]. The idea
consists in considering the more general case allowing for nonsimply
connected handles on the Riemann surface; i.e. we shall also consider
handles on the Riemann surface which, under cutting, give rise to
pairs of discs that are no longer disconnected to each other [19]. This
would ultimately imply partial or total breakdown of the Schottky
group invariance under projective transformations between discs [21].
Such handles need not be
on shell in the sense that the analogue of the Wheeler DeWitt operator
acting on the excited eigenstates $\Psi_{m_{n}^{(i)}}(Y_{n}^{(i)})$ is no
longer zero, but gives the corresponding harmonic-oscillator eigenvalues
$nm_{n}^{(i)}$ [19,20]. In this case, the $\delta$ function for all field modes
in the Green function should be replaced [19] by a path integral which
has the $x_{n}$ fixed at the given values at time $t=0$ on a circle
and $t=t_{1}$ on another, for some Euclidean time interval $t_{1}$
between the two circles. Unlike the pure-state case, this path
integral is not generally separable into a product of wave
functions (4), but, after integrating over time $t_{1}$, it gives
the density matrix for a mixed quantum state,
and is equal to the
propagator
\begin{equation}
K(x_{n},0;x'_{n},t_{1})=\prod_{n>0,i}\sum_{m_{n}^{(i)}>0}\Phi_{m_{n}^{(i)}}(Y_{n}^{(i)})\Phi_{m_{n}^{(i)}}(Y_{n}^{(i)'})e^{-nm_{n}^{(i)}t_{1}},
\end{equation}
where $\Phi_{m_{n}^{(i)}}(Y_{n}^{(i)})$ would match the excited
eigenstates of the harmonic oscillators ($\Phi_{m_{n}^{(i)}}$
$\equiv\Psi_{m_{n}^{(i)}}$) if we allowed
an unlimited resolution for the fields, or equal some wave functions which
contain only that part of the information contained in the $\Psi_{m_{n}^{(i)}}$
that associates with the finite eigenenergy sector surviving
after the introduction of a given cut off at a finite highest
energy scale. Therefore, we are interpreting here the loss
of coherence contained in the density matrix as the information loss
induced by a physical cut off at finite energy.
If the handles are nonsimply connected, and hence off shell, then,
instead of the quantum constraint equation (3),
we must use the "time-independent" wave
equation [19,20]
\begin{equation}
(H_{WDW}-\sum_{n>0,i}nm_{n}^{(i)})\Psi=0,
\end{equation}
so that the wave function for the wormhole handles becomes
\begin{equation}
\Psi\equiv\Psi(x_{0},Y)=e^{iKx_{0}}\prod_{n>0,i}e^{-\frac{1}{2}n(Y_{n}^{(i)})^{2}}.
\end{equation}

The $K^{2}$ term in (6), resulting from the application of operator
$-\frac{\partial^{2}}{\partial x_{0}^{2}}$ to $\Psi$,
gives the energy of the $x_{0}$ plane waves, with
$K^{2}<0$. This would correspond to a timelike momentum in {\it Lorentzian}
space-time. Wick rotating to Euclidean momentum, $K\rightarrow -iK_{E}$,
we have from (6) and (7)
$K_{E}^{2}=\sum_{n>0,i}(m_{n}^{(i)}+\frac{1}{2})n$
and, since we are
dealing with harmonic oscillators, we have in general
$K_{E}=\sum_{n>0,i}m_{n}^{(i)}n$ [22].
Let us then calculate the quantity
\[\sum_{m_{n}^{(i)}>0}\Psi(x_{0},Y)\Psi^{*}(x_{0}',Y')\]
\begin{equation}
=\sum_{m_{n}^{(i)}>0}\prod_{n>0,i}(e^{-\frac{1}{2}n(Y_{n}^{(i)})^{2}}e^{-\frac{1}{2}n(Y_{n}^{(i)'})^{2}}e^{-m_{n}^{(i)}n(x_{0}'-x_{0})}).
\end{equation}
The relative minus sign between $x_{0}$ and $x_{0}'$ in (8) should be kept
anyway in order to ensure an orientable surface when the handles are glued
together [14]. Taking $x_{0}'-x_{0}=t_{1}$ and noting that, since each
two circles can have any time separations, one should integrate (8) over
all possible values of $t_{1}$ [19], we can see that
\[i\int dt_{1}K(x_{n},0;x_{n}',t_{1})\]
\begin{equation}
=\sum_{m_{n}^{(i)}>0}\int d(x_{0}'-x_{0})\Psi(x_{0},Y)\Psi^{*}(x_{0}',Y')\equiv
i\rho(Y;Y'),
\end{equation}
whenever we take for the states $\Phi_{m_{n}^{(i)}}$ in the propagator (5) only
that
part of the $\Psi_{m_{n}^{(i)}}$ which corresponds to the harmonic oscillator
ground states, surviving after projecting off all the information
contained in the Hermite polynomials. This is the effect on the quantum
state from having a finite maximum resolution in our string model, or
equivalently a Virasoro condition $K^{2}=2$ [16]. We shall see later on
that this
condition is obtained as a pole when integrating the Green function
for the wave function over the radius of the circles in the
handle dynamics.

{}From (7) and (9), the density matrix for nonsimply connected handles becomes
\begin{equation}
\rho(Y;Y')=\sum_{m_{n}^{(i)}>0}\prod_{n>0,i}\frac{\Psi_{0}(Y_{n}^{(i)})\Psi_{0}(Y_{n}^{(i)'})}{nm_{n}^{(i)}},
\end{equation}
where $\Psi_{0}(Y_{n}^{(i)})=e^{-\frac{1}{2}n(Y_{n}^{(i)})^{2}}$.
Thus, for each $m_{n}^{(i)}>0$ of
nonsimply connected handles, we should use a Green function for
the density matrix of the mixed state case given by

\[<x_{\mu}(z_{1}).x_{\nu}(z_{2})...>\]
\begin{equation}
=\int
d[x_{\mu}]x_{\mu}(z_{1})x_{\nu}(z_{2})...\prod_{r}\prod_{n>0,i}\tilde{\Theta}(x_{0}-x_{0}')\Psi_{0}(Y_{n}^{(i)})\Psi_{0}(Y_{n}^{(i)'})e^{-I},
\end{equation}
where we have replaced the full $\delta$ function in (1) for
the density matrix (7), specialising to a single generic
relative probability $(nm_{n}^{(i)})^{-1}$ for each mode
$n>0$, and the step function
$i\tilde{\Theta}(x_{0}-x_{0}')=\Theta(x_{0}-x_{0}')$ arises from
integrating the $\delta$ function for the zero mode over its
argument, as it is done in (9) and (10). In the present
approach, if we want to consider a Green function also for
the pure state case, instead of the full $\delta$ function,
we should use the probability $\mid\Psi\mid^{2}$ obtained
from (7) as the weighting factor. This Green function will
then be
\[<x_{\mu}(z_{1}).x_{\nu}(z_{2})...>\]
\begin{equation}
=\int
d[x_{\mu}]x_{\mu}(z_{1})x_{\nu}(z_{2})...\prod_{r}\prod_{n>0,i}\delta(x_{0}-x_{0}')\Psi_{0}(Y_{n}^{(i)})\Psi_{0}(Y_{n}^{(i)'})e^{-I}.
\end{equation}

If we regard each pair of circles as the ends of a sum of wormholes of
different species [12,13], then each species would now be labelled by just the
momentum
$K$ of the zero mode, but {\it not} the levels $m_{n}^{(i)}$ of the other
modes. The
quantity $K$ can be interpreted as the conserved scalar charge carried
by the wormhole [16], and the
lack of any spectral dependence of the wave function (7) should express
the feature that the wormholes {\it have no quantum hair} [23].
This requirement can alternatively be regarded as either
expressing the existence of a finite maximum resolution at
$K^{2}$, or a boundary condition on the quantum state of
stringy wormholes. In any case, it is associated with a
breakdown of the Schottky invariance under projective
transformations [21] for the $n>0$ modes. One would regard
this breaking
mechanism as the string analogue of a breakdown of diffeomorphism
invariance in 4-dimensional spacetime [24,25]. It extends also to the zero
modes
for nonsimply connected handles whose quantum state is given
by a mixed statistical density matrix. In this case,
associated to each state
$\Psi_{0}(Y_{n}^{(i)})$, there are infinitely many relative
probabilities $\frac{1}{nm_{n}^{(i)}}$ [19].

\section{STRINGY WORMHOLE DYNAMICS}

We can now calculate the effect of wormholes on tachyonic amplitudes,
for nonsimply connected handles, whose quantum state is given by both
a density matrix and a wave function,
using the
procedure devised by Lyons and Hawking [14].
For the case of handles in mixed state
and tachyons with momenta $p_{j}$,
unlike the pure-state case considered in Ref. 14,
the path integral describing the interaction cannot be
factorised into path integrals on each of the two circles [19].
Instead of the wave function (4), one should then introduce
as weighting factor the density matrix element
$\rho_{m_{n}}^{(i)}$ (see below)
which corresponds to each relative probability
$\frac{1}{nm_{n}^{(i)}}$. In the limit of small circle radius
$r\rightarrow 0$, we then have for each of these density
matrix elements
\[D(\rho;p_{j})\propto\int
dx_{0}dx_{0}'(\prod_{n>0}dY_{n}^{(i)}dY_{n}^{(i)'})\rho_{m_{n}^{(i)}}\]
\[\times\int drr^{-3+(\sum_{1}^{M}p_{j})^{2}}\int[{\it
D}\zeta_{j}]\exp[i(x_{0}-x_{0}')\sum_{1}^{M}p_{j}]\]
\begin{equation}
\times\prod_{n>0}e^{\sum_{i}[-\frac{1}{2}n((Y_{n}^{(i)})^{2}+(Y_{n}^{(i)'})^{2})+ir^{n}k_{n}^{(i)}(Y_{n}^{(i)}-Y_{n}^{(i)'})]},
\end{equation}
where
\[\rho_{m_{n}^{(i)}}=\prod_{n>0}\frac{\Psi_{0}(Y_{n}^{(i)})\Psi_{0}(Y_{n}^{(i)'})}{nm_{n}^{(i)}},\]
and we take, as in [14],
$r=\mid z_{2}\mid^{-1}$, $\zeta_{j}=\frac{z_{j}}{z_{2}}$, $j=3,...,M$,
with $M$ the number of on-shell tachyon vertex operators inserted in the
region exterior to the circles, $\zeta_{0}=0$, $\zeta_{1}=\infty$,
$\zeta_{2}=1$. $\int[{\it D}\zeta_{j}]$ denotes integration over $\zeta_{j}$
with a measure whose explicit form need not be known for our calculation,
and $k_{n}=2\sum_{j=1}^{M}p_{j}\zeta_{j}^{-n}$, with $k_{n}^{(1)}$
and $k_{n}^{(2)}$ the real and imaginary parts of $k_{n}$, respectively.

Each integral pair over $Y_{n}^{(i)}$ and $Y_{n}^{(i)}$ gives a
factor $\frac{e^{-\frac{r^{2n}(k_{n}^{(i)})^{2}}{n}}}{n}$ for each $n$.
The Gaussian exponential factor would only contribute higher-order
interactions and will be disregarded in our calculation [14]. The other
possible contribution would come from integration over each pair of zero-mode
fields $x_{0}$,$x_{0}'$. Since all possible dependence of the density
matrix on such field has already been integrated out, unlike for handles
in a pure state,
we are left with
a single delta on $\sum_{j=1}^{M}p_{j}$, so in first order approximation
the path integral (13) gives essentially a factor
$\int\frac{dr}{r^{3}m_{n}^{(i)}n^{2}}$
for each $m_{n}^{(i)}$ and $n$.
Note that this factor contains no integration over momentum
$K$. As in the space-time wormhole case [15], the effect of
handles will be given again by a vertex operator, located
approximately at the center of the circles [12,13].
Thus, for each $m_{n}^{(i)}$ and $n$,
the stringy wormholes in mixed state will give rise
to a bi-local effective action [12] for each $m_{n}^{(i)}$
\begin{equation}

--Boundary (ID u4ldalpcRaJ1Kt26Lirckg)
Content-type: TEXT/PLAIN; CHARSET=ISO-8859-1
d\sigma_{2}\sum_{q,i}\frac{V_{q}(\sigma_{1})V_{q}(\sigma_{2})}{m_{n}^{(i)}n^{2}},
\end{equation}
where $V_{q}$ are the vertex functions. Following hereafter the same
procedure as for wormholes in space-time [15], or simply
connected handles on a Riemann surface [19],
this action can be made local
by introducing $\alpha$ parameters, i.e.
\begin{equation}
\int

d\sigma\sum_{q,i}((-\frac{1}{2}\alpha_{q}^{2}m_{n}^{(i)}n^{2})+\alpha_{q}V_{q}(\sigma))
\end{equation}
which leads to a probability measure over the $\alpha$ parameters
for each $m_{n}^{(i)}$
\begin{equation}
Z(\alpha)\prod_{q,i}e^{-\frac{1}{2}\alpha_{q}^{2}m_{n}^{(i)}n^{2}},
\end{equation}
where again
$Z(\alpha)$ is the path integral over all fields $x^{\mu}$ on the
two-sphere [12-14],
containing the effective interaction $\alpha_{q}V_{q}(\sigma)$.
The distribution for $\alpha$-parameters associated with (16)
corresponds to just one of the infinite relative probabilities
for the state $\Psi_{0}(Y_{n}^{(i)})$ of handles. Therefore,
unlike in the pure-state case, one should sum (16) over all
$m_{n}^{(i)}$ [8,26], to finally obtain a probability measure
\begin{equation}
Z(\alpha)\prod_{q}(e^{\frac{1}{2}n^{2}\alpha_{q}^{2}}-1)^{-1}
\end{equation}
for each $n$ and $i$. Thus, like for nonsimply connected wormholes
in 4-dimensional space-time [8,26],
we obtain a Planckian distribution for $\alpha$
parameters that allows to interpret $\frac{1}{2}\alpha_{q}^{2}$ as the
energy of the quanta of a radiation field, and $n^{-2}$ as some temperature,
on string-theory superspace.

In the case that the quantum state of the handles be given by the
wave function (7),
the calculation is similar, but with
$\rho_{m_{n}^{(i)}}$ replaced by $\Psi(x_{0},Y)\Psi^{*}(x_{0}',Y')$
in the path integral (13). In actual calculation,
the only difference is in the integration
over the field zero modes which now produces $\delta(K-\sum_{j=1}^{M}p_{j})$.
The essential factor becomes then $-\frac{1}{n(K^{2}-2)}$ for each $n$.
The pole in this factor will consistently
give the expected Virasoro string result $K^{2}=2$ [16].
It follows that each mode $n$ contributes a probability measure
\begin{equation}
Z(\alpha)\prod_{q,i}e^{-\frac{1}{2}\alpha_{q}^{2}(1-\frac{K^{2}}{2})n}=Z(\alpha)\prod_{q,i}e^{-\frac{1}{2}\alpha_{q}(K)^{2}n^{2}},
\end{equation}
where $\alpha_{q}(K)^{2}=\alpha_{q}^{2}\frac{(1-\frac{K^{2}}{2})}{n}$.
Similarly to as it was done for nonsimply connected wormholes
in space-time [8,26],
this Gaussian law should be interpreted as a "Wien law" for
the radiation field on the infinite dimensional space of all
coordinate field values on a circle (string-theory superspace),
derivable from (17) in the limit of small exponent. These
results indicate that the actual  probability distribution
for the effective coupling constants $\alpha$ shows an
essential quantum discontinuity for the $\alpha$ which is
over and above that is implied by the Coleman Gaussian law.

Note moreover that, since the $\alpha$ parameters
in both (17) and (18) are labelled by the momentum $K$, but {\it not} the
levels $m_{n}^{(i)}$, our results do not allow any
interpretation of the $\alpha$ in terms of
a quantum field on superspace, which is
dimensionally reducible to an
infinite tower of fields on space-time. For handles in a pure state where,
if the initial state is a state with definite values of the $\alpha$
parameters, the final state will be the same as the initial state [27],
according to our results, the radiation field $\alpha$ can
be dimensionally reduced to just one field, rather than a
tower of fields, on the usual space-time, i.e. on the prefered
minisuperspace from string-theory superspace, consisting of
just the $n=0$ modes [12,13]. Therefore, there will always exist
a set of classical values for $\alpha$ which makes it possible
to drive a consistent mechanism that fixes the values of
the coupling constants. In the pure state case, for such a
mechanism we can take the Coleman mechanism for wormholes
in space-time [1,4] because, unlike originally thought [3],
there will moreover be a prefered physical cut off which is
implied by the occurrence of the natural finite resolution
in string theory, discussed in this paper, that prevents
divergences in the probability distribution.
The existence of such a physical cut off offers also a possible explanation
to why all the infinite extra dimensions (other than the field
zero modes on a circle which would make usual space-time) of string-theory
superspace are not noticed by us [12,13].
We note finally that the Coleman mechanism
that makes $\Lambda=0$, could not fix the values of all other
coupling constants if the handles are in mixed states, since in this
case quantum coherence can no longer be preserved [28]. For
in such a situation, however, physical constants could still
be fixed by the alternative statistical mechanism which leads
to the identification of perturbations to the effective action
with minus the entropy of the system [8].

\vspace{1.3cm}

{\bf Acknowledgements}

This work was supported by CAICYT under Research Project N§ PB91-0052.
The author wish to thank the hospitality of Department of Applied
Mathematics and Theoretical Physics, Cambridge Univ., UK,
where part of this work was done.

\pagebreak

\noindent\section*{References}
\begin{description}
\item [1] S. Coleman, Nucl. Phys. B307, 867 (1988).
\item [2] I. Klebanov, L. Susskind and T. Banks, Nucl. Phys. B317, 665 (1989).
\item [3] S.W. Hawking, Nucl. Phys. B335, 155 (1990).
\item [4] J. Preskill, Nucl. Phys. B323, 141 (1989).
\item [5] G.W. Gibbons, S.W. Hawking and M.J. Perry, Nucl. Phys. B138, 141
(1978).
\item [6] W. Fischler and L. Susskind, Phys. Lett. B217, 48 (1989).
\item [7] J. Polchinski, Phys. Lett. B219, 251 (1989).
\item [8] P.F. Gonz\'alez-D\'{\i}az, Mod. Phys. Lett. A8, 1089 (1993).
\item [9] S. Coleman and K. Lee, Phys. Lett. B221, 242 (1989).
\item [10] P.F. Gonz\'alez-D\'{\i}az, Phys. Lett. B233, 85 (1989).
\item [11] P.F. Gonz\'alez-D\'{\i}az, Int. J. Mod. Phys. A7, 2355 (1992).
\item [12] S.W. Hawking, Nucl. Phys. B363, 117 (1991).
\item [13] S.W. Hawking, Phys. Script. T36, 222 (1991).
\item [14] A. Lyons and S.W. Hawking, Phys. Rev. D44, 3802 (1991).
\item [15] S.W. Hawking, Phys. Rev. D37, 904 (1988).
\item [16] A. Lyons, Ph. D. Thesis, University of Cambridge, UK, 1991.
\item [17] M. Karliner, I. Klebanov and L. Susskind, Int. J. Mod. Phys. A3,
1981 (1988).
\item [18] D. Amati, M. Ciafaloni, and G. Veneziano, Phys. Lett. B216, 41
(1989).
\item [19] P.F. Gonz\'alez-D\'{\i}az, Nucl. Phys. B351, 767 (1991).
\item [20] S.W. Hawking, in {\it 300 Years of Gravitation}, eds. S.W. Hawking
and W. Israel (Cambridge Univ. Press, Cambridge, 1987).
\item [21] S. Mandelstam, in {\it Unified Field Theories, Proceedings of
the 1985 Santa Barbara Workshop}, eds. M. Green and D. Gross (World
Scientific, Singapore, 1986).
\item [22] W.H. Louisell, {\it Radiation and Noise in Quantum Electronics}
(McGraw Hill, New York, 1964).
\item [23] S. Coleman, J. Preskill and F. Wilczek, Phys. Rev. Lett.
67, 1975 (1991) (see also references therein) have discussed the
possibility that black holes may carry gauge quantum hair. This
was questioned by considering the effect of wormholes on
quantum coherence in P.F. Gonz\'alez-D\'{\i}az, Mod. Phys. Lett.
A8, 2539 (1993).
\item [24] S.B. Giddings, Phys. Lett. B268 (1991).
\item [25] P.F. Gonz\'alez-D\'{\i}az, Phys. Lett. B307, 362 (1993).
\item [26] P.F. Gonz\'alez-D\'{\i}az, Class. Quant. Grav. 10, 2505 (1993).
\item [27] S. Coleman, Nucl. Phys. B310, 643 (1988).
\item [28] P.F. Gonz\'alez-D\'{\i}az, Phys. Rev. D45, 499 (1992).

\end{description}

\end{document}